\documentclass[11pt]{article}
\usepackage{amsmath,amssymb,color,epsfig,cite}
\usepackage{graphicx}
\usepackage{setspace}
\usepackage{mathrsfs}
\usepackage[english]{babel}
\usepackage{inputenc}

\usepackage{amsfonts}

\makeatletter
\@addtoreset{equation}{section}
\makeatother

\newcommand{\be}{\begin{equation}}
\newcommand{\ee}{\end{equation}}
\newcommand{\bea}{\setlength\arraycolsep{2pt} \begin{eqnarray}}
\newcommand{\eea}{\end{eqnarray}}

\def\0{{\sst{(0)}}}
\def\1{{\sst{(1)}}}
\def\2{{\sst{(2)}}}
\def\3{{\sst{(3)}}}
\def\4{{\sst{(4)}}}
\def\5{{\sst{(5)}}}
\def\6{{\sst{(6)}}}
\def\7{{\sst{(7)}}}
\def\8{{\sst{(8)}}}
\def\sst#1{{\scriptscriptstyle #1}}

\begin{document}

\vspace{15pt}
\begin{center}
{\Large {\bf A Note on the Near Horizon Charges for the Five Dimensional Myers-Perry Black Holes} }

\vspace{15pt}
{\bf Zahra Mirzaiyan}

\vspace{10pt}

 {\it Max-Planck-Institut f\"ur Gravitationsphysik (Albert-Einstein-Institut), \\
M\"uhlenberg 1, D-14476 Potsdam, Germany.}

\vspace{10pt}


\vspace{20pt}

\underline{ABSTRACT}
\end{center}
Inspired by the recent work on the spacetime structure near generic black hole horizons \cite{Grumiller:2019fmp}, the near horizon charges for an explicit example in higher dimensions than four $(d>4)$, namely for the five dimensional Myers-Perry metric with two equal rotation parameter are found in Hamiltonian formalism.  Finding the supertranslation and the one-form superrotation, it is proved that the Myers-Perry black hole with  two equal rotation parameter $a=b$ does not satisfy the gauge flatness condition due to the non-vanishing associated field strength in five dimensional spacetime. It is shown that as the near horizon limit of such a metric satisfies a specific set of boundary conditions, the near horizon algebra can be represented as an infinitely many copies of Heisenberg algebras as a generalisation to the Kerr case in four dimensions.
\noindent

\thispagestyle{empty}

\vfill
zahra.mirzaiyan@aei.mpg.de

\pagebreak
\tableofcontents 
\section{Introduction}
The current note is based on \cite{Grumiller:2019fmp} which mainly shows that any non-extremal\footnote{The metric can be brought into a Rindler form.} (finite temperature) horizon has an infinite set of near horizon symmetries and associated soft hair excitations in the sense of Hawking, Perry and Strominger \cite{HPS}. Defining suitable near horizon boundary conditions while working in Hamiltonian formalism (providing the near horizon conditions for spatial part of metric and its canonical momentum) \cite{Dirac,Arnowitt,Regge}, integrable finite near horizon charges (associated to non-trivial diffeomorphisms) are calculated. It is also shown that for horizons that are either flat or non-rotating, the near horizon symmetry can be represented as infinitely many copies of Heisenberg algebras in any spacetime dimensions. In this note, we consider an explicit example in higher dimensions than four. Near horizon supertranslation and superrotation charges are calculated and it is shown that five dimensional Myers-Perry black hole with $a=b$ does satisfy  the results presented in \cite{Grumiller:2019fmp} as an example in higher dimensions. However, we consider only a very maximally symmetric case for the Myers-Perry black holes, which is the case with two equal rotation parameter. We only focus to find the charges here rather than near horizon algebra as the near horizon metric obviously satisfy the boundary conditions in \cite{Grumiller:2019fmp} and therefore, the near horizon algebra consists of infinitely many copies of the Heisenberg algebra which extends the results from lower dimensions \cite{Afshar1,Afshar2}. We start with the known Kerr case and re-derive the near horizon Heisenberg-like generators which we call them supertranslation and superrotation charges. We then generalise the method to the five dimensional case by finding the Heisenberg-like near horizon generators for the Myers-Perry black holes with $a=b$ by presenting a detailed calculation in Hamiltonian formulation of General Relativity.

\section{Setup}
The metric at the near horizon of a regular metric with the surface gravity $\kappa$ can be written in a Rindler-like form as ($a$ and $b$ shows the angular coordinates and runs from one in case of a three dimensional black hole to higher dimensions)
\begin{eqnarray}
ds^2=-\kappa \rho^2 dt^2+d\rho^2+\Omega_{ab} dx^a dx^b+...,
\end{eqnarray}
where, $\Omega_{ab}$ is the metric transverse to the horizon\footnote{The determinant of $\Omega$ is non-zero to guarantee a non-singular metric on the horizon. Therefore, the Tylor expansion in the near horizon region is allowed as we will use such expansions in the calculations.}. In the Hamiltonian formulation of GR the metric can be written in the following form ($i,j=\text{radial coordinate and angular coordinates}$)
\begin{eqnarray}
ds^2=-N^2 dt^2+g_{ij} (dx^i+N^i dt)(dx^j+N^j dt),
\end{eqnarray}
where, $N$ is the laps function, $N^i$s are shift functions and $g_{ij}$ is the spatial part of the metric. One needs to provide near horizon conditions for the spatial part of metric\footnote{These conditions are defined in the boundary conditions (II.2) of \cite{Grumiller:2019fmp}. However, we only need the conditions for $g_{ij}$ and the defined boundary conditions in (\ref{ex}).}. The associated canonical momenta $\pi^{ij}$ is found as
\begin{eqnarray}\label{ex}
&&\pi^{\rho a}=\frac{\Pi^{\rho a}}{16\pi G}+\mathcal{O}(\rho^2), \ \ \pi^{\rho \rho}=\mathcal{O}(\rho), \ \ \pi^{ab}=\mathcal{O}(\rho),\nonumber\\
&& N=\kappa \rho +\mathcal{O}(\rho^3),\ \ N^\rho=\mathcal{O}(\rho^3), \ \  N^a=\mathcal{O}(\rho^2).
\end{eqnarray}
The above near horizon boundary conditions are preserved by a set of diffeomorphisms generated by a set of vector field $\xi^\mu$ ($\mu=0,1,2,...$) described by the vector field introduced in (II.3) of \cite{Grumiller:2019fmp}. It can be shown that the canonical charges can be defined as
\begin{eqnarray}
\mathcal{P}:=\frac{\sqrt{\Omega}}{8\pi G},\ \ \mathcal{J}_a:=\Omega_{ab} \frac{\Pi^{\rho b}}{8\pi G \sqrt{\Omega}},
\end{eqnarray}
where, $\mathcal{P}$ and $\mathcal{J}_a$ are near horizon supertranslations and superrotations. $\mathcal{P}$ is a scalar and $\mathcal{J}_a $ is a one-form that can be decomposed into an exact, coexact and a harmonic part\footnote{Refer to \cite{Afshar:2018apx} for detailed discussion on the decomposition of $p$-forms.}. It is shown that if $\mathcal{J}_a$ is locally exact $\mathcal{J}_a=8\pi G \partial_a \mathcal{Q}$, which means the associated field strength $F_{ab}^H:=(d \mathcal{J}^H)_{ab}$ is zero.\\
Then the following Poisson bracket construct the Heisenberg algebra as
\begin{eqnarray}\label{algebra}
\{ \mathcal{Q}(x), \mathcal{P}(y)\}=\frac{1}{8\pi G} \delta (x-y).
\end{eqnarray}
We try to present the Heisenberg-like generators  in Hamiltonian formalism, $\mathcal{P}$ and $\mathcal{J}_a$ for the Kerr metric in details which is consistent with the calculations in \cite{Grumiller:2019fmp}. Then, we generalise this calculations to the simplest possible rotating black hole in higher dimensions, five dimensional Myers-Perry black hole with two equal angular momenta.

\section{Kerr black holes}

The Kerr metric in Boyer-Lindquist coordinates $(t,r,\theta, \phi)$ is as fallows
\begin{eqnarray}\label{kerrmetric}
ds^2=-dt^2 +\frac{2M r}{\Sigma} \left(a\ \text{sin}^2 \theta \ d\phi -dt \right)^2 +\Sigma \left(\frac{dr^2}{\Delta}+d\theta^2 \right) +\left(r^2 +a^2\right) \text{sin}^2 \theta \ d\phi^2,
\end{eqnarray}
with 
\begin{eqnarray}\label{detsig}
\Delta:= r^2 -2M r +a^2, \ \ \ \ \ \Sigma:= r^2+a^2 \text{cos}^2 \theta.
\end{eqnarray}
To find the location of Killing horizons, one has to find roots of $\Delta=0$ where we have infinite red shift as
\begin{eqnarray}
\Delta=0 \ \ \Longrightarrow \ \ \ r_\pm = M \pm \sqrt{M^2 -a^2}.
\end{eqnarray}
Introducing the new parameter $R=r_- / r_+$, mass, rotation parameter and the surface gravity can be found as

\begin{eqnarray}\label{mak}
&&M=\frac{r_+ +r_-}{2}=\frac{r_+}{2} (1+R),\ \ \ a^2 =r_+ r_- = r_+^2 R,\ \ \ \kappa=\frac{r_+}{a^2 +r_+^2}-\frac{1}{2 r_+}=\frac{1-R}{2 r_+(1+R)}.\nonumber\\
\end{eqnarray}
Now we should look for a coordinate transformation such that shifts the outer horizon of Kerr black hole $r=r_+$ to $\rho=0$ and brings it to a form where the near horizon metric has a Rindler form. Choosing the following coordinate transformations with introducing the parameter $\alpha(\theta)=\frac{1-R}{2 r_+ (1+R\text{cos}^2 \theta)} $,
\begin{eqnarray}\label{cochange}
&&r= r_{+}+ \frac{\alpha(\theta)}{2}\ \rho^2, \ \ \phi=\varphi+\frac{2\kappa \sqrt{R}}{1-R} t.
\end{eqnarray}
Therefore, the boosted Kerr metric in the coordinates $(t,\rho, \theta, \varphi)$ reads as

\begin{eqnarray}\label{bmetric}
ds^2&=&\left(-1 +\frac{8M r}{\Sigma} \frac{\kappa^2 R}{(1-R)^2} a^2 \text{sin}^4 \theta +\frac{2M r}{\Sigma} -\frac{8M r}{\Sigma} \frac{\kappa \sqrt{R}}{1-R} \ a \ \text{sin}^2 \theta+ (r^2 +a^2) \frac{4 \kappa^2 R \text{sin}^2 \theta }{(1-R)^2}\right) d\tau^2\nonumber\\
&&+\left(\frac{8M r}{\Sigma} \frac{\kappa \sqrt{R}}{1-R} a^2 \text{sin}^4 \theta-\frac{4M r}{\Sigma} a\ \text{sin}^2 \theta +(r^2 +a^2) \frac{4\kappa \sqrt{R}}{1-R} \text{sin}^2 \theta \right) d\tau d \varphi\nonumber\\
&&+\left(\frac{\Sigma}{\Delta} \alpha^2 \ \rho^2 \right)\ d\rho^2+\left(\frac{\Sigma}{\Delta} \alpha \alpha^{\prime}\rho^3\right) d\rho \ d\theta+\left( \Sigma+\frac{\Sigma}{\Delta}\frac{{\alpha^{\prime}}^2}{4} \rho^4\right) d\theta^2\nonumber\\
&&+ \left(\frac{2M r}{\Sigma} a^2 \text{sin}^4 \theta +(r^2 +a^2) \ \text{sin}^2 \theta \right) d\varphi^2.
\end{eqnarray}
We are interested to know how the Kerr metric looks like in the near horizon region at $\rho\rightarrow0$. Using the following expansions in this limit,
\begin{eqnarray}\label{rsigmadelta}
&&\frac{r}{\Sigma}=\frac{1}{r_{+} \ (1+R\ \text{cos}^2 \theta)}+\mathcal{O}(\rho^2),\nonumber\\
&&\frac{\Sigma}{\Delta}=\frac{2 r_{+} \ (1+R\ \text{cos}^2 \theta) }{(1-R) \ \alpha \ \rho^2}+\frac{1-2R-R\ \text{cos}^2 \theta}{(1-R)^2}+\frac{(R-3) \ \alpha \ \rho^2}{2 r_{+} \ (1-R)^2}+\mathcal{O}(\rho^4),
\end{eqnarray}
the near horizon expansion of the components of the metric (\ref{bmetric}) reads as
\begin{eqnarray}\label{tt}
&&g_{tt}=-\frac{(1-R)^2}{4 r_{+}^2 \ (1+R)^2}\rho^2 +\mathcal{O} (\rho^4),\nonumber\\
&&g_{t\varphi}=-\frac{(1-R) \ \sqrt{R} \ (-6 +(-3+R)\ R -(1-R) \ R \ \text{cos} [2\theta]) \ \text{sin}^2 \theta}{8 (1+R) r_{+} (1+R\ \text{cos}^2 \theta)^2 }\rho^2+\mathcal{O} (\rho^4),\nonumber\\
&&g_{\rho\rho}=1+\mathcal{O}(\rho^2),\nonumber\\
&&g_{\rho \theta} =\frac{R\ \text{sin}\theta \ \text{cos}\theta \ \ }{ (1+R \ \text{cos}^2 \theta)} \ \rho +\mathcal{O}(\rho^3),\nonumber\\
&&g_{\theta \theta}= r_+^2 (1+R \ \text{cos}^2 \theta) +\mathcal{O}(\rho^2),\nonumber\\
&&g_{\varphi \varphi}=\frac{ r_+^2 \ \text{sin}^2 \theta \ (1+R)^2 }{(1+R\ \text{cos}^2 \theta)}
+\mathcal{O} (\rho^2).\nonumber\\
\end{eqnarray}
\subsection{Three-plus-one dimensional decomposition of the Kerr metric in the near horizon region}

In the context of Arnowitt-Deser-Misner (ADM) decomposition, the Laps and shift functions, and the spatial part of the metric can be obtained by (Note that $i,j=\rho, \theta, \phi$)

\begin{eqnarray}\label{34}
{}^3 g_{ij}\equiv{}^4 g_{ij}, \ \ \ \ \ \ N_i\equiv{}^4 g_{0i}, \ \ \ \ N\equiv (-{}^4 g^{tt})^{(-1/2)}.
\end{eqnarray}
The conjugate momenta can be found by
\begin{eqnarray}
\pi^{ij}=\sqrt{-{}^4g} \ ({}^4\Gamma_{pq}^0 -g_{pq} \ {}^4\Gamma_{rs}^0 \ g^{rs} ) \ g^{ip}\ g^{jq}.
\end{eqnarray}
Thus, one can easily find
\begin{eqnarray}\label{33}
&&N_\varphi=-\frac{(1-R) \ \sqrt{R} \ (-6 +(-3+R)\ R -(1-R) \ R \ \text{cos} [2\theta]) \ \text{sin}^2 \theta}{8 (1+R) r_{+} (1+R\ \text{cos}^2 \theta)^2 }\rho^2+\mathcal{O} (\rho^4),\nonumber\\
&&N_\rho=N_\theta= 0,\nonumber\\
&&N=\kappa \ \rho +\mathcal{O}(\rho^3).\nonumber\\
\end{eqnarray}
The inverse of the spatial part of the near horizon metric in three dimensions, ${}^3 g^{ij}$ reads as
\begin{eqnarray}
{}^3 g^{ij}={}^4 g^{ij}+(N^i \ N^j/N^2).
\end{eqnarray}
Therefore,
\begin{eqnarray}
&&{}^3 g^{\rho \rho}=1+\mathcal{O} (\rho^2),\nonumber\\
&&{}^3 g^{\rho \theta}=-\frac{R \ \text{sin}\theta \ \text{cos}\theta}{r_{+}^2 \ (1+R \ \text{cos}^2 \theta)^2} \rho+\mathcal{O} (\rho^3),\nonumber\\
&&{}^3 g^{\theta \theta}=\frac{1}{r_{+}^2 \ (1+R \ \text{cos}^2 \theta)}+\mathcal{O} (\rho^2),\nonumber\\
&&{}^3 g^{\varphi \varphi}=\frac{1+R \ \text{cos}^2 \theta}{r_{+}^2 \ (1+R)^2 \ \text{sin}^2 \theta}+\mathcal{O} (\rho^2).\nonumber\\
\end{eqnarray}

Since $\sqrt{-{}^4 g}=N \sqrt{{}^3 g}$ we have ${}^3g=r_{+}^4 (1+R)^2 \ \text{sin}^2 \theta$, where $g$ is the determinant of the metric.
The associated momenta reads as
\begin{eqnarray}
&&\pi^{\rho \varphi}=-\frac{\sqrt{R} \ \text{sin}\theta \ (3+R+(R-R^2) \ \text{cos}^2 \theta) }{(1+R) \ (1+R \ \text{cos}^2 \theta)}+\mathcal{O}(\rho^2),\nonumber\\
&&\pi^{\theta \varphi}=\mathcal{O}(\rho),\nonumber\\
&&\pi^{\rho \rho}=\pi^{\theta \theta}=\pi^{\varphi \varphi}=0.\nonumber\\
\end{eqnarray}
The horizon metric is obtained as
\[
\Omega_{ab}=\left({\begin{array}{cc}
r_+^2 (1+R\ \text{cos}^2 \theta)&0\\
0&r_+^2 \frac{ \ \text{sin}^2 \theta \ (1+R)^2 }{1+R\ \text{cos}^2 \theta}\\

\end{array}}\right).
\]
with the determinant
\begin{eqnarray}\label{deter}
\Omega=r_+^4 \text{sin}^2 \theta \ (1+R)^2.
\end{eqnarray}
The state-dependent Heisenberg-like generators $\mathcal{P}$ and $\mathcal{J}_a^H$ thus, read as 
\begin{eqnarray}\label{p}
&&\mathcal{P}=\frac{\sqrt{\Omega}}{8\pi G}=\frac{r_+^2 \ (1+R)}{8\pi G}\  \text{sin} \theta,\\
&&\mathcal{J}_a^H=\delta_{a}^{\varphi} \Omega_{\varphi \varphi} \frac{\pi^{\rho \varphi}}{8\pi G \sqrt{\Omega}}=-\delta_{a}^{\varphi} \sqrt{R} \frac{3+R+(R-R^2)\text{cos}^2 \theta}{ \ 8\pi G \ (1+R \text{cos}^2 \theta)^2 }\text{sin}^2 \theta.
\end{eqnarray}
The associated field strength for the Kerr black holes reads as 
\begin{eqnarray}
F_{\theta\varphi}^H=\frac{\sqrt{R}(1+R)^2 (R\text{cos}^2\theta-3)\ \text{sin}[2\theta]}{(1+R\text{cos}^2\theta)^3}.
\end{eqnarray}
Therefore, due to the non-vanishing value of $F_{\theta\varphi}^H$, the Kerr black hole does not satisfy the gauge flatness condition $F_{ab}^H=0$. However, the flux through the horizon associated with $F_{\theta\varphi}^H$ is zero due to a regular horizon\footnote{For the case with NUT charges, the flux through the horizon is non-zero.}. As noted in \cite{Grumiller:2019fmp}, the superrotation generator $\mathcal{J}_a^H$ has a coexact part with together with $\mathcal{P}$ makes the charge algebra denoted in (\ref{algebra}). In the next section, we generalise a similar calculation for the five dimensional case.

\section{Five dimensional Myers-Perry black holes with $a=b$}

The five dimensional Myers-Perry metric reads as 

\begin{eqnarray}\label{MP5}
&&ds^2= -dt^2 +\frac{M}{\Sigma} \left(dt-a\ \text{sin}^2 \theta \ d\phi -b\ \text{cos}^2 \theta \ d\psi \right)^2 \nonumber\\
&&+ \Sigma \left(\frac{r^2}{\Delta} dr^2 +d\theta^2 \right)+(r^2 +a^2) \ \text{sin}^2 \theta \ d\phi^2 +(r^2 +b^2)\ \text{cos}^2 \theta \ d\psi^2, 
\end{eqnarray}
where, $M$ is the mass and $a$ and $b$ are rotation parameters. Assuming a very special case with $a=b$, $\Sigma$, $\Delta$ and the inner and outer horizons read as

\begin{eqnarray}\label{pisigab}
&&\Sigma=r^2 +a^2, \ \ 
\Delta=(r^2 +a^2) \ (r^2 +a^2)- M r^2,\nonumber\\
&&r_{\pm}^2 = \frac{1}{2} \left(M-2a^2 \pm \sqrt{M^2 -4 M \ a^2 }\right).
\end{eqnarray}
Introducing a new parameter $R=r_-/r_+$, the rotation parameter, mass and the surface gravity can be read as
\begin{eqnarray}\label{mak}
\noindent{a^2=r_+^2 R},\ \
M=r_+^2 (1+R)^2, \ \
\kappa=\frac{1-R}{r_+ (1+R)}.
\end{eqnarray}
Using the following coordinate change with $\gamma=\frac{\sqrt{R}}{r_{+} \ (1+R)}$ and $\beta=\frac{1-R}{2 r_{+}}$ brings in the Rindler form and  shifts the outer horizon $r=r_+$ to $\rho=0$.
\begin{eqnarray}\label{cochange}
r=r_{+} +\beta (\theta) \ \rho^2,\ \ 
\phi=\varphi+ \gamma (\theta) \ t, \ \
\psi=\chi+ \gamma (\theta) \ t.
\end{eqnarray}
Using the above coordinate transformations, the components of the metric (\ref{MP5}) in $(t,\rho,\theta,\varphi,\chi)$ coordinates in the near horizon region are as follows

\begin{eqnarray}
&&g_{tt}=-\frac{(1-R)^2}{r_{+}^2 \ (1+R)^2} \rho^2 +\frac{(1-R)^2 \ (3+R^2)}{4 \ r_{+}^4 \ (1+R)^3}\rho^4 +\mathcal{O}(\rho^6),\nonumber\\
&&g_{\rho\rho}=1+\frac{(3-4R+3R^2)}{4\ r_{+}^2 \ (1+R)}\rho^2+\frac{(2R^3+R^2-3)}{2 r_{+}^4 \ (1+R)^2}\rho^4+\mathcal{O}(\rho^6),\nonumber\\
&&g_{\theta \theta}=r_+^2 (1+R) +(1-R)\rho^2+\frac{(1-R)^2}{4 r_{+}^2}+\mathcal{O}(\rho^6),\nonumber\\
&&g_{\varphi\varphi}=r_+^2 (1+R)\ \text{sin}^2 \theta \ (1+R \text{sin}^2 \theta )+\frac{1}{2} (1-R)\ (2-R+R\ \text{cos}[2\theta])\ \text{sin}^2 \theta \rho^2 \nonumber\\
&&+\frac{(1-R)^2 \ (2+(R-5)\ R +(R-3) \ R\ \text{cos}[2\theta]) \text{sin}^2 \theta}{8 \ r_{+}^2 \ (1+R)}\rho^4+\mathcal{O}(\rho^6),\nonumber\\
&&g_{\chi\chi}=r_+^2 (1+R) \ \text{cos}^2 \theta \ (1+R \text{cos}^2 \theta )-\frac{1}{2} (1-R) (-2+R+R\ \text{cos}[2\theta]) \ \text{cos}^2 \theta \rho^2\nonumber\\
&&-\frac{(1-R)^2 \ (-2-(R^2-5R) +(R-3) \ R\ \text{cos}[2\theta])\ \text{cos}^2 \theta}{8 \ r_{+}^2 \ (1+R)}\rho^4+\mathcal{O}(\rho^6),\nonumber\\
&&g_{\varphi\chi}=R r_+^2 \ (1+R) \ \text{sin}^2 \theta \ \text{cos}^2 \theta -(1-R) \ R \ \text{sin}^2 \theta \ \text{cos}^2 \theta \ \rho^2\nonumber\\
&& -\frac{(R-3) \ (1-R)^2 \ R \ \text{sin}^2 [2\theta]}{16 \ r_{+}^2 \ (1+R)}\rho^4+\mathcal{O}(\rho^6),\nonumber\\
&&g_{\varphi t}=\frac{2 \ (1-R) \ \sqrt{R} \ \text{sin}^2 \theta}{r_{+} \ (1+R)}\rho^2+\frac{(1-R)^3 \ \sqrt{R} \ \text{sin}^2 \theta}{2 \ r_{+}^3 (1+R)^2}\rho^4+\mathcal{O}(\rho^6),\nonumber\\
&&g_{\chi t}=\frac{2 \ (1-R)\ \sqrt{R} \ \text{cos}^2 \theta}{r_{+} \ (1+R)}\rho^2+\frac{(1-R)^3 \ \sqrt{R} \ \text{cos}^2 \theta}{2 \ r_{+}^3 \ (1+R)^2}\rho^4+\mathcal{O}(\rho^6),\nonumber\\
&&g_{\rho\theta}=g_{\varphi\theta}=g_{\chi\theta}=0.
\end{eqnarray}

We can find the metric on the horizon as
\begin{eqnarray}\label{horizonmet}
&&\Omega_{ab} dx^a dx^b =r_+^2 [ (1+R)\ \text{sin}^2 \theta \ (1+R \text{sin}^2 \theta ) \ d\varphi^2 +2 R \text{sin}^2 \theta \ \text{cos}^2 \theta d\varphi d\chi \nonumber\\
&& + (1+R) \ \text{cos}^2 \theta \ (1+R \text{cos}^2 \theta ) d\chi^2 + (1+R) d\theta^2 ],
\end{eqnarray}
with the determinant
\begin{eqnarray}\label{det}
\Omega=r_+^6 (1+R)^4\ \text{sin}^2 \theta \ \text{cos}^2 \theta.
\end{eqnarray}
Thus, using Eq. (\ref{det}), the near horizon supertranslation charge can be found as
\begin{eqnarray}\label{p}
\mathcal{P}=\frac{\sqrt{\Omega}}{8\pi G}=\frac{r_+^3 (1+R)^2}{8\pi G}\ \text{sin} \theta \ \text{cos} \theta.
\end{eqnarray}
The horizon metric (\ref{horizonmet}) is topologically a 3-sphere with Ricci scalar
\begin{eqnarray}\label{ricci}
\mathcal{R}=\frac{2 (\mathcal{R} -3)}{r_+^2 (\mathcal{R}+1)}.
\end{eqnarray}
The Euler characteristic is

\begin{eqnarray}\label{euler}
\int_{0}^{2\pi} d\varphi \ \int_{0}^{2\pi} d\chi \ \int_{0}^{\pi} d\theta \sqrt{\Omega} \ \mathcal{R}=0.
\end{eqnarray}

\subsection{Four-plus-one dimensional decomposition of the metric}
To find the near horizon superrotation generators, we need the conjugate momenta. Therefore, we do a four plus one decomposition in the context of ADM decomposition. The spatial components of the metric, the laps and shift functions together with the conjugate momenta can be obtained using (Note that $i,j=\rho, \theta, \varphi, \chi$)

\begin{eqnarray}\label{34}
&&{}^4 g_{ij}\equiv{}^5 g_{ij}, \ \ \ \ \ \ N_i\equiv{}^5 g_{0i}, \ \ \ \ N\equiv (-{}^5 g^{tt})^{(-1/2)},\nonumber\\
&&\pi^{ij}=\sqrt{-{}^5g} \ ({}^5\Gamma_{pq}^0 -g_{pq} \ {}^5\Gamma_{rs}^0 \ g^{rs} ) \ g^{ip}\ g^{jq}.
\end{eqnarray}
Thus, the laps function reads as
\begin{eqnarray}\label{laps}
N= (-{}^5 g^{tt})^{(-1/2)}= \kappa \rho^2 \ (1+\mathcal{O}(\rho^2).
\end{eqnarray}
Since $N^i=N^2 \ {}^{5} g^{ti}$, one can find the shift functions as 

\begin{eqnarray}\label{nphi}
&&N^{\varphi}=\frac{2 \ (1-R) \ \sqrt{R}}{r_{+}^3\ (1+R)^3 }\rho^2+\mathcal{O}(\rho^4),\\
&&N^{\chi}=\frac{2 \ (1-R) \ \sqrt{R}}{r_{+}^3\ (1+R)^3 }\rho^2+\mathcal{O}(\rho^4),\\
&&N^\rho=N^\theta=0.
\end{eqnarray}
The spatial components of the metric ${}^4g^{ij}$ can be obtained as
\begin{eqnarray}
{}^4 g^{ij}={}^5 g^{ij}+(N^i \ N^j/N^2).
\end{eqnarray}
Thus,
\begin{eqnarray}
&&{}^4 g^{\rho\rho}=1+\mathcal{O}(\rho^2),\nonumber\\
&&{}^4 g^{\theta\theta}=\frac{1}{r_{+}^2 \ (1+R)}-\frac{(1-R)}{r_{+}^4 \ (1+R)^2}\rho^2+\mathcal{O}(\rho^4),\nonumber\\
&&{}^4 g^{\varphi\chi}=-\frac{R}{r_{+}^2 \ (1+R)^2}+\mathcal{O}(\rho^2),\nonumber\\
&&{}^4 g^{\varphi\varphi}=\frac{ \ (1+R \ \text{cos}^2 \theta)}{\ r_{+}^2 \ (1+R)^2 \ \text{sin}^2 \theta}+\mathcal{O}(\rho^2),\nonumber\\
&&{}^4 g^{\chi\chi}=\frac{ \ (1+R \ \text{sin}^2 \theta)}{\ r_{+}^2 \ (1+R)^2 \ \text{cos}^2 \theta}+\mathcal{O}(\rho^2).\nonumber\\
\end{eqnarray}

The conjugate momenta can be found by
\begin{eqnarray}\label{con}
&&\pi^{\rho \varphi}=2 \ \sqrt{R} \ \text{sin}\theta \ \text{cos}\theta+\mathcal{O}(\rho^ 2).\\
&&\pi^{\rho \chi}=2 \ \sqrt{R} \ \text{sin}\theta \ \text{cos}\theta+\mathcal{O}(\rho^ 2).\label{con2}
\end{eqnarray}
Using the associated conjugate momenta mentioned in (\ref{con}) and (\ref{con2}), the superrotation generators
 $\mathcal{J}_a^H=\Omega_{ab}\frac{\pi^{\rho b}}{8\pi G \sqrt{\Omega}}$, can be found as

\begin{eqnarray}
&&\mathcal{J}_\varphi^H=2\ \sqrt{R} \ \text{sin}^2 \theta,\\
&&\mathcal{J}_\chi^H=2\ \sqrt{R} \ \text{cos}^2 \theta.
\end{eqnarray}
Thus, the one-form superrotaion reads as
\begin{eqnarray}\label{oneform}
\mathcal{J}^H=2 \ \sqrt{R}\ (\text{sin}^2 \theta \ d\varphi+ \text{cos}^2 \theta \ d\chi).
\end{eqnarray}
Myers-Perry black hole does not satisfy the gauge flatness condition due to non-vanishing associated field strength as
\begin{eqnarray}
F_{\theta \varphi}^H=(d \mathcal{J}^H)_{\theta \varphi}=- F_{\theta \chi}^H=2 \ \sqrt{R} \ \text{sin}[2\theta].
\end{eqnarray}
However, similar to the Kerr case, the flux through the 3-sphere horizon associated with $F_{ab}^H$ is zero. The coexact part of the Heisenberg-like generator (\ref{oneform}) together with the supertranslation charge (\ref{p}), construct the charge algebra introduced in \cite{Grumiller:2019fmp} as the near horizon boundary conditions are satisfied. It should be noted that, here we only considered a very maximally symmetric Myers-Perry black hole with two equal rotation parameter which is an explicit example of metric which satisfies the boundary conditions defined in \cite{Grumiller:2019fmp} in higher dimensions than four.

\section*{Acknowledgements} 
We are grateful of Daniel Grumiller, Behrouz Mirza, Hamideh Nadi, Mahsa Salimi and especially Aditya Mehra for the comments on the draft. We also acknowledge the Erwin Schr\"{o}dinger Institute where this work started through JRF support.


\begin{thebibliography}{99}

\bibitem{Grumiller:2019fmp} 
  D.~Grumiller, A.~Pérez, M.~M.~Sheikh-Jabbari, R.~Troncoso and C.~Zwikel,
  ``Spacetime structure near generic horizons and soft hair,''
  Phys.\ Rev.\ Lett.\  {\bf 124}, no. 4, 041601 (2020), arXiv:1908.09833 [hep-th].
  
  \bibitem{HPS}
  S. W. Hawking, M. J. Perry, and A. Strominger, Phys.
Rev. Lett. \textbf{116}, 231301 (2016), arXiv:1601.00921 [hep-th].
  
  \bibitem{Dirac}
P. A. M. Dirac, Rev. Mod. Phys. \textbf{21}, 392 (1949).

\bibitem{Arnowitt}
R. L. Arnowitt, S. Deser, and C. W. Misner, Gen. Rel.
Grav. \textbf{40}, 1997 (2008), arXiv:gr-qc/0405109 [gr-qc].

\bibitem{Regge}
T. Regge and C. Teitelboim, Ann. Phys. \textbf{88}, 286 (1974).

\bibitem{Afshar1}
H. Afshar, S. Detournay, D. Grumiller, W. Merbis,
A. Perez, D. Tempo, and R. Troncoso, Phys. Rev. D\textbf{93},
101503 (2016), arXiv:1603.04824 [hep-th].

\bibitem{Afshar2}
H. Afshar, D. Grumiller, W. Merbis, A. Perez, D. Tempo,
and R. Troncoso, Phys. Rev. D\textbf{95}, 106005 (2017),
arXiv:1611.09783 [hep-th].

\bibitem{Afshar:2018apx} 
  H.~Afshar, E.~Esmaeili and M.~M.~Sheikh-Jabbari,
  ``Asymptotic Symmetries in $p$-Form Theories,''
  JHEP {\bf 1805}, 042 (2018),arXiv:1801.07752 [hep-th].
 
\end{thebibliography}
\end{document}